**Signature of checkerboard fluctuations in the phonon spectra of a possible polaronic metal La$_{1.2}$Sr$_{1.8}$Mn$_2$O$_7$.**

F. Weber[1,2,*], N. Aliouane[3,**], H. Zheng[4], J.F. Mitchell[4], D.N. Argyriou[3], and D. Reznik[1,5]

[1] Forschungszentrum Karlsruhe, Institut für Festkörperphysik, POB 3640, D-76021 Karlsruhe, Germany
[2] Physikalisches Institut, Universität Karlsruhe, D-76128 Karlsruhe, Germany
[3] Helmholtz-Zentrum Berlin für Materialen und Energy, Glienicker Str. 100, D-14109 Berlin, Germany.
[4] Materials Science Division, Argonne National Laboratory, Argonne, Illinois, 60439, USA
[5] Laboratoire Léon Brillouin, CEA Saclay, F-91191 Gif-sur-Yvette, France

*present address: Materials Science Division, Argonne National Laboratory, Argonne, Illinois, 60439, USA
**present address: Institute For Energy Technology, P.O. Box 40, NO-2027 Kjeller, Norway

**Charge carriers in low-doped semiconductors may distort the atomic lattice and trap themselves forming so-called small polarons[1,2]. High carrier concentrations can lead to short range ordered polarons (large polarons)[3,4] and even to long range charge and orbital order[5]. Both systems should be insulating with a large electrical resistivity, which decreases with increasing temperature[6]. However, photoemission measurements recently found a polaronic pseudogap, in a metallic phase of La$_{2-2x}$Sr$_{1+2x}$Mn$_2$O$_7$[7]. This layered manganite is famous for colossal magnetoresistance (CMR) associated with a phase transition from this low-temperature metallic phase to a high temperature insulating phase.[7,8,9] Broad charge order peaks due to large polarons observed by neutron and x-ray scattering in the insulating phase disappear when La$_{2-2x}$Sr$_{1+2x}$Mn$_2$O$_7$ becomes metallic[10]. We report results of inelastic neutron scattering measurements showing that polarons remain inside the metallic phase as fluctuations that strongly broaden and soften certain phonons near the wave vectors where the charge order peaks appeared in the insulating phase. Our findings imply that polaronic signatures in metals may generally come from a competing insulating charge-ordered phase. It is highly relevant to cuprate superconductors with both a pseudogap[11,12] and a similar phonon effect associated with a competing stripe order[13].**

The competition between a ferromagnetic metallic (FM) and paramagnetic insulating (PM) phase is behind the famous CMR phenomenon in manganite perovskites such as the quasicubic La$_{1-x}$Ca$_x$MnO$_3$ and its bi-layer counterpart La$_{2-2x}$Sr$_{1+2x}$Mn$_2$O$_7$. CMR occurs close to a metal-insulator transition temperature, $T_c$, when applied magnetic field, which favors the FM phase, raises $T_c$. This phase competition appears due to natural coupling between magnetism and orbital ordering as Goodenough demonstrated in the 1950's in LaMnO$_3$[14]. In simple terms the $e_g$ orbitals of Jahn-Teller active Mn$^{3+}$ form a staggered arrangement with the ferromagnetic exchange in MnO$_2$ layers and antiferromagnetic inter-planar coupling. Doping this orbitally ordered Mott insulator with $x$ holes leads to other arrangements where orbital, charge and spin order strongly interact. One such example is the charge and orbitally ordered insulator (known as CE-type[15]) at half doped compositions such as La$_{0.5}$Ca$_{0.5}$MnO$_3$[16] or LaSr$_2$Mn$_2$O$_7$[17,18,19] where the $e_g$ charge is localized on alternate Mn sites producing to first order distinct Mn$^{3+}$ and Mn$^{4+}$ sites[16,18,20] (see fig. 1(b)). The Jahn-Teller effect and orbital ordering of the $e_g$-charge induces a characteristic distortion with a superlattice peak at a wave vector of $\mathbf{q}_{CE}=(1/4,1/4,0)$ in LaSr$_2$Mn$_2$O$_7$. For intermediate compositions the frustration of the charge and orbital ordering results in a ferromagnetic metallic ground state, with CMR in the vicinity of $T_c$.[5] A phase transition to an insulating phase occurs at elevated temperatures where 15Å clusters of CE-order[3] mingle with charge-delocalized clusters[21] as clearly visible diffuse elastic or quasielastic diffraction peaks in the vicinity of $\mathbf{q}_{CE}$[3,22]. Below $T_c$ ferromagnetic fluctuations quickly melt the local CE correlations[3,18]. Instability of these polarons in the presence of ferromagnetism and applied magnetic field leads to CMR close to $T_c$. This ferromagnetic metal has poor conductivity and small carrier scattering length[9].

While the disappearance of the polaronic scattering below $T_c$ was thought to mark the onset of a conventional double-exchange metallic ground state, recent photoemission measurements of the bi-layer manganite La$_{2-2x}$Sr$_{1+2x}$Mn$_2$O$_7$ with $x=0.4$[7] show an unconventional pseudogap in many respects similar to features predicted to arise from single polarons[23,24]. Their results seem to imply a coexistence between polarons and metallic electrical resistivity, which defines a polaronic metal whose nature remains enigmatic. We present results of inelastic neutron scattering measurements suggesting that polaronic effects in the bi-layer manganites are dominated by strong fluctuations that mimic the extended lattice distortions of the CE type that emerge above $T_c$. Their experimental signature is phonon broadening and softening at the wave vectors where diffuse scattering due to short range CE order appears above $T_c$. Our findings show that the polaronic degrees of freedom dominate the properties of the ferromagnetic metallic phase of this CMR manganite and reveal that in the strong coupling limit a polaronic metal is a viable ground state.

We focused on the Mn–O bond stretching phonon branches dispersing in the [110] direction in the bilayer CMR compound, $La_{1.2}Sr_{1.8}Mn_2O_7$. $MnO_6$ octahedra in this manganite align along the three crystallographic axes (fig. 1(a)). Lattice distortions in the CE polaronic state have the ordering wavevector $q_{CE}$ = (1/4, 1/4, 0) (Fig. 1b. The lattice deformation around the nominal $Mn^{3+}$ site (indicated by the occupied $e_g$ orbital) corresponds to a part of the eigenvector of the transverse phonon (fig. 1(c)), whereas the deformation around the nominal $Mn^{4+}$ site corresponds to a part of the eigenvector of the longitudinal one (fig. 1(d)). This partial matching provides a natural mechanism of strong coupling between fluctuating CE-distortions and both the transverse and the longitudinal phonons at the same wave vector.

In contrast, in the commonly studied "113" quasicubic compounds such as $La_{0.7}Sr_{0.3}MnO_3$ tilted octahedra result in extra bond-bending branches folded in from different parts of the Brillouin zone. These mix with the stretching vibrations away from the zone center[25,26] prohibiting the study of pure bond-stretching modes. Previous studies of quasicubic manganites have not been able to link very strong temperature-induced changes in phonon intensities[27,28] with specific charge or orbital fluctuations. Thus the bilayer manganites are ideal for investigating dispersions of bond-stretching modes.

Our sample was a high-quality 0.5cm$^3$ single-crystal of $La_{1.2}Sr_{1.8}Mn_2O_7$ with the FM transition temperature measured at 115 K. The experiments were carried out on the 1T triple-axis spectrometer at the ORPHEE reactor, using doubly focusing Cu220 monochromator and PG002 analyzer crystals. Final energies were fixed at 30.5 meV or 35 meV. The components ($Q_x$, $Q_y$, 0) are expressed in reciprocal lattice units (rlu) (1 rlu = $2\pi/a$, a = 3.88 Å). Measurements were performed in the constant-momentum-transfer **Q** mode, **Q** = **τ** + **q**, and **τ** is a reciprocal lattice point. The longitudinal/transverse branch was measured at **q**=(-h, -h, 0) / (-h, +h, 0) (0≤h≤0.5) respectively.

Small bond-stretching phonon structure factors were overcome by careful selection of the experimental conditions, coverage of two Brillouin zones (BZ), and very long counting times of as much as 1 hour per point at 10K and 2 hrs at 150K. We assigned the observed phonon peaks by comparing our results both to shell model predictions and to the previous measurements of layered cuprate perovskites, $La_{2-x}Sr_xCuO_4$.[29] The shell model has been developed for the cuprates and has already been successfully applied to a cubic manganite.[25] It predicts phonon dispersions reasonably well with the notable exception of the longitudinal bond-stretching branch. This is also true for the bilayer manganites. Additional special force constants can adjust calculated dispersions of the stretching branches to agree with the measured ones. After the adjustment, the model makes good qualitative predictions of phonon eigenvectors, and, therefore, intensities, which we had used to identify bond-bending and bond-stretching peaks (fig. 2). The peak assignment procedure and background determination are described in detail in the supplementary online material. The resolution function was calculated for our experimental setup and folded with the dispersions of the bond stretching branches in order to obtain the intrinsic phonon linewidths shown in fig. 3 (b) and (d).

Figure 2 shows representative scans between 40 and 80 meV measured deep inside the FM phase (at 10K) in the BZ adjacent to **τ** = (3, 3, 0). The zone center bond-stretching vibration at **Q** = (3, 3, 0) is at 73 meV (fig. 2(a)). There is no peak corresponding to the 43 meV bond-bending vibration in figure 2(a) because its structure factor is zero. We measured its dispersion around **Q** = (4, 4, 0) where its structure factor is strong. Another bond bending peak, at 33meV, is just below the energy range of figure 2(a).

The transverse bond-stretching branch direction disperses steeply downwards along (h,-h,0) and the transverse bending branches disperse gradually upwards. They do not overlap until h=0.4 allowing a study of pure bond-stretching modes at (0≤h≤0.35). They can be traced as a high energy shoulder of the much stronger bending modes further towards the zone boundary (h=0.5).

The longitudinal bond-stretching branch disperses downwards along (h,h,0), whereas the higher-energy bending mode disperses sharply upwards with an anticrossing near h=0.3. However, the structure factor for this bending mode remains near zero in the **Q**=(3, 3, 0) BZ up to close to the zone boundary (h=0.5), thus the observed intensity unambiguously comes from the stretching mode at ($0 \leq h \leq 0.4$). (see supplementary online material for details)

Both the transverse and longitudinal stretching branches dip sharply and have a linewidth maximum (shortest lifetime) near $q_{CE}$ (fig. 3), which indicates strong electron-phonon coupling deep in the metallic state. In $La_{1.2}Sr_{1.8}Mn_2O_7$ $q_{CE}$ corresponds to the maximum of the diffuse scattering at 128K (Fig. 3b,d) due to static (frozen) short-range CE order[18]. This observation highlights the close relationship between the polaronic insulating phase and the phonon renormalization in the metallic phase. The q-width of the phonon broadening is somewhat larger than of the elastic peak. It corresponds to a correlation length of 2 - 3 unit cells compared with 4 unit cells in the insulating phase.

This phonon anomaly is absent from the equivalent branches in the cuprates.[29] It also appears to be absent from the quasicubic $La_{0.7}Sr_{0.3}MnO_3$ where the downward dispersion of the transverse branch was reported to continue to lower energies at low temperatures.[25] This difference may be explained by the fact that no CE order appears in this system in the paramagnetic state and that the continual softening towards the zone boundary may result from interaction with uncorrelated small polarons.

The longitudinal branch disperses downwards (fig. 3c), whereas the shell model suggests an upward dispersion. This discrepancy is similar to overdoped $La_{1.7}Sr_{0.3}CuO_4$. The overall downward dispersion most probably reflects metallic behavior as opposed to an exotic effect[29]. Where the difference between $La_{1.2}Sr_{1.8}Mn_2O_7$ and the cuprates is dramatic, is near $q_{CE}$: In the former there is a sharp dispersion dip, accompanied by linewidth broadening, whereas in the latter it follows a cosine dispersion with a roughy constant linewidth.

We now discuss what happens above $T_c$ when the static short-range CE order appears. Much longer counting times are required here because static distortions suppress coherent phonon intensities by almost a factor of 2 while the background increases by a factor of 1.5. For this reason we measured only one wave vector at **Q**=(2.75, 3.25, 0) at 150K, well above the static polaron ordering temperature.

Figure 4 shows the background-subtracted phonon spectra at **Q** = (2.75, 3.25, 0) measured at 10 K and 150 K. Because the contribution of the lower energy phonons cannot be satisfactorily evaluated we fitted the data with a single Gaussian above 56 meV. In this energy range the observed intensity is of pure bond stretching character (as follows from Fig. 2d). We see no indication of anharmonic broadening and softening usually expected at increased temperatures as the amplitude of atomic motion increases.

Broadening due to static disorder is expected to increase above $T_c$ due to the formation of CE-polarons. Since such an increase does not occur for the bond-stretching phonons, their broadening and softening at low temperatures must be of electronic origin. We note that this behavior is in contrast with the bond-bending phonon at 43meV, which does broaden above $T_c$ (see Fig. S3 in supplementary online material). The question, whether the temperature dependence in figure 4 originates from the increased availability of electronic states in the low temperature phase or from the enhanced charge-phonon coupling, cannot be answered at this point.

We conclude that phonon renormalization in the [110] direction is dominated by a fluctuation of CE-type, which we assign to CE dynamic polarons. This fluctuation exists below $T_c$ and then condenses into the static or quasistatic order[18] at the metal-insulator transition. This novel phenomenon is unusual, because polaron fluctuations have been observed at $T<0.1T_c$ and therefore do not appear to be associated with the critical behavior near the phase transition. The polaron fluctuation around $q_{CE}$ = (1/4, 1/4, 0) may be related to the strong nesting of the Fermi surface[1] near the same wave vector[7,8]. More work is necessary to clarify the association between these two phenomena.

Our measurements offer the defining characteristics of a *polaronic metal*.[7] It combines quasiparticles that exist over a small portion of the Fermi surface with CE-type modes inherited from the polaronic insulator above $T_c$, which strongly renormalize certain phonons. Further, charge transport below $T_c$ is dominated by coherent polarons whose wave functions overlap, in contrast to the classical single polarons found at small carrier concentrations. Indeed above $T_c$ such overlap leads to static local CE-order. One cannot avoid the striking similarities between the [110] direction in this manganite and the [100] direction in the superconducting cuprates where charge fluctuations in the form of stripes seem to induce a similar phonon renormalization. It is intriguing that real and imaginary parts of the low temperature phonon self-energies in the cuprates at $\mathbf{q} \approx (¼, 0, 0)$ and in the manganites at $\mathbf{q} \approx (¼, ¼, 0)$ are very close. This similarity indicates that dynamic charge stripes in the cuprates are somehow analogous to the fluctuating checkerboard-like CE-polarons in the manganites. Exploring this connection will no doubt shed light on many unresolved questions for both classes of compounds where pseudogap-like phases play a critical role.

**Acknowledgement:** The authors would like to thank M. Braden and L. Pintschovius for helpful discussions and suggestions, and M. Braden and W. Reichardt for allowing us to use shell model parameters for $La_{2-2x}Sr_{1+2x}Mn_2O_7$ that they found in a separate investigation. Work at Argonne National Laboratory was supported under Contract No. DE-AC02-06CH11357 by UChicago Argonne, LLC, Operator of Argonne National Laboratory, a U.S. Department of Energy Office of Science Laboratory.

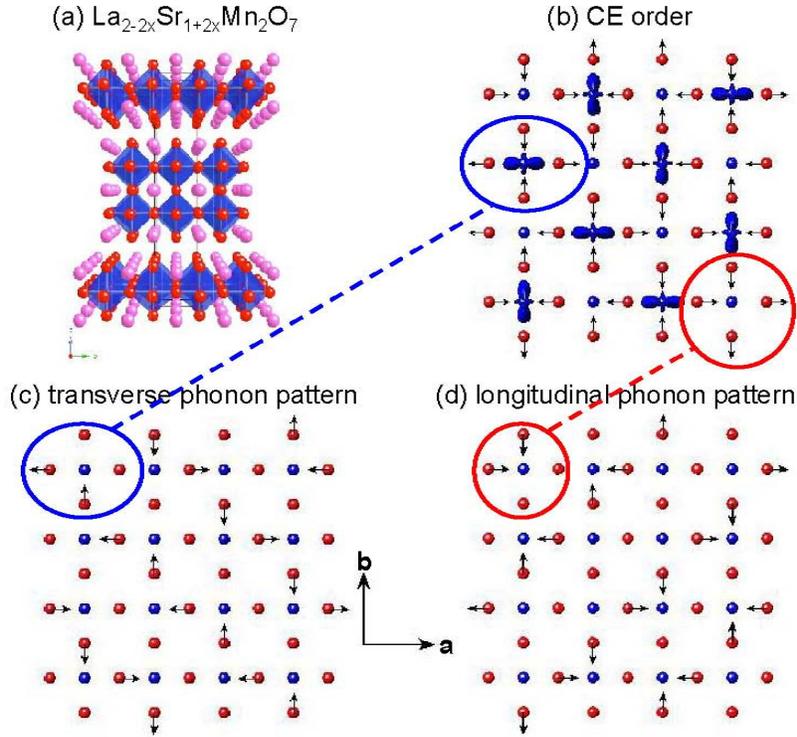

**Figure 1: Correspondence between CE-order and phonon vibration patterns.** (a) Layered crystal structure of $La_{2-2x}Sr_{1+2x}Mn_2O_7$ with La (purple) atoms between the $MnO_6$ octahedra (Mn: blue, O: red). (b) Schematic of the displacements of oxygens from the ideal unordered structure in long-range CE ordered half-doped compositions (x=0.5). Drawn $e_g$ orbitals indicate the $Mn^{3+}$ ions. Mn displacements are not shown. (c),(d) Eigenvectors of the transverse (c) and longitudinal (d) bond-stretching phonons with $\mathbf{q}$ = (0.25, 0.25, 0). Small Mn displacements are not shown. Short range CE fluctuations should couple strongly to both the longitudinal and transverse phonons. The dashed lines indicate the partial matching of oxygen displacements of the CE order and of the phonon eigenvectors.

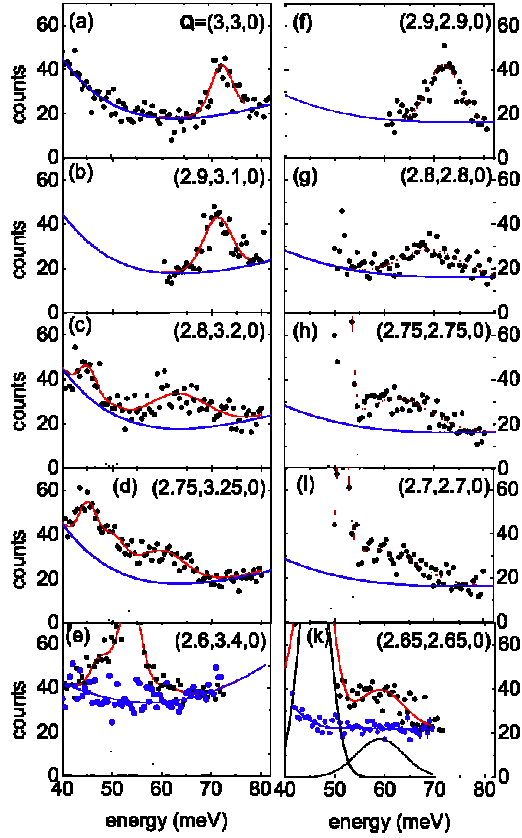

**Figure 2: Representative const-Q scans for transverse and longitudinal phonons dispersing in the (1, 1, 0) direction.** Data are normalised to a monitor of 50 000 neutron counts, which corresponds to a counting time of 5 min. for an energy transfer E = 70 meV. Scans shown in panels (e) and (k) are each measured in different runs and therefore, have different backgrounds. See supporting material for details about the background determination. The scan at **Q** = (2.6, 3.4, 0) was measured with a final energy $E_f$ = 30.5 meV. All other data were measured with $E_f$ = 35 meV. Functions fitted to the background data are shown as blue lines. Black lines are Gaussians fitted to the data after background subtraction. Red lines represent the background function plus the Gaussian fits. Data were taken at T = 10 K. Error bars represent s.d.

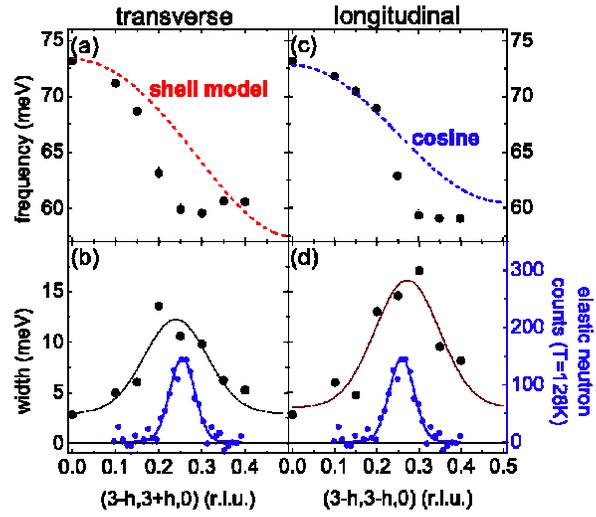

**Figure 3: Dispersions and linewidths of transverse and longitudinal bond-stretching phonons in the (1, 1, 0) direction of $La_{1.2}Sr_{1.8}Mn_2O_7$.** (a),(c) Data points represent measured phonon energies; black lines are guides to the eye; red dashed line is the dispersion calculated by the shell model. (b),(d) Black dots are phonon FWHM after correction for the experimental resolution. Black lines are guides to the eye. Blue dots represent elastic intensity (right hand scale) of the short-range charge and orbital order above $T_c$ = 115 K. Error bars represent s.d.

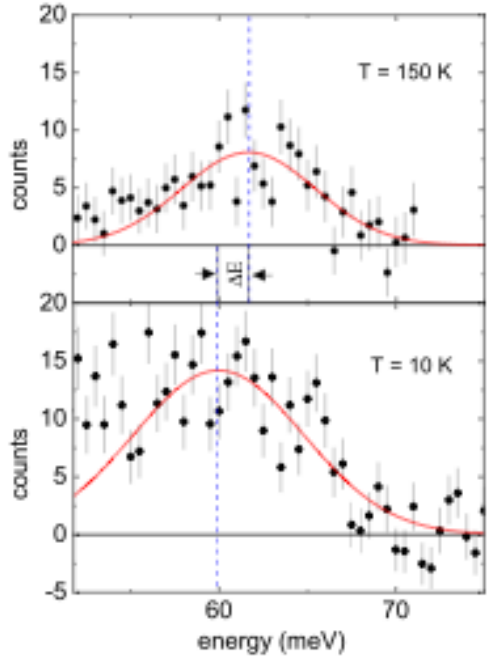

**Figure 4: Background subtracted phonon spectra at Q = (2.75, 3.25, 0) at 150 K (top) and 10 K (bottom).** The background was measured separately at the two temperatures. Red lines are Gaussians fitted to the observed peaks for E > 56 meV (see text). The vertical dashed lines mark the fitted frequencies of the bond stretching phonon at T = 10 K and 150 K.

# Supporting Material: Background determination and peak assignment

In order to determine the properties of the Mn-O bond stretching branches in $La_{1.2}Sr_{1.8}Mn_2O_7$ it was important to separate it from other modes that may be close in energy or even hybridise with it. We took advantage of character specific intensity variations of phonons with different eigenvectors by measuring in two Brillouin zones (BZ), i.e. around the zone center points of (3, 3, 0) and (4, 4, 0). These results were compared to the shell model for the lattice dynamics of $La_{1.2}Sr_{1.8}Mn_2O_7$, which was developed by W. Reichardt and M. Braden in a separate investigation. We found that the model correctly reproduced many features of phonon dispersions mostly correctly, although there were some differences between the model and the data for the observed intensities of bond-bending phonons. Nevertheless, we believe that our detailed measurements combined with the shell model prediction allow determination of the bond-stretching phonon dispersions in the bilayer manganite $La_{1.2}Sr_{1.8}Mn_2O_7$.

## 2. Background Determination

In order to extract the phonon signal correctly, it is necessary to determine the background that typically originates from incoherent scattering, multiphonon scattering and other processes that do not have strong dependence on either wave vector or energy. After performing many different scans almost completely covering two BZ we found that the background is nearly wavevector independent in a large part of the energy range of interest. This section explains how the background was determined.

The background intensity was obtained from data taken at energies away from phonons. In general, we used the zone center scan to obtained the background for $38 \leq E \leq 70$ meV. On the high energy end we used data points of scans in which the bond stretching phonon has already dispersed to relatively low energies, i.e. $\mathbf{q} \geq (0.25, 0.25, 0)$.

In scans around (3, 3, 0) we found that transverse scans (including the zone center) show an upturn in the background at high energies, which is not present in the longitudinal scans (figs. 2 and S1). Therefore, the background functions shown in fig. 2 differ at high energies in the left and right column. In fig. S1 we show longitudinal scans taken in the (3, 3, 0) and (4, -4, 0) BZ. Apparently, the background for longitudinal scans is equal for both BZ for 50 meV $\leq E \leq$ 70meV. Although we have no data to evaluate a difference in the background for longitudinal and transverse scans around (3, 3, 0) below 50 meV, equivalent data around (4, -4, 0) show a similar less curved, i.e. lower, background for longitudinal scans at low energies. In the end we used the background function shown in figs. 2 (f)-(k) and S1 in the analysis for all longitudinal scans. We note that the uncertainty in the background for the longitudinal scans in the (3, 3, 0) zone below 50 meV has very little quantitative and no qualitative effect on our data analysis.

3. Zone Center

The zone center scans in the two Brillouin zones are shown in fig. S1. According to the shell model as well as general experience[25-28] with measurements on perovskites, the upper peak at 73 meV can be unambiguously assigned to the bond-stretching branch. The peak at 43 meV, which appears only at (4, 4, 0) belongs to the upper bending branch. Intensities of the two peaks agree with the shell model predictions.

**4. Transverse Dispersion**

We show our data for the (3, 3, 0) BZ in the letter. We also determined the dispersion of the transverse bond-bending modes, by making scans in the BZ around (4, -4, 0) for $\mathbf{Q} = (4+x, -4+x, 0)$, with $x = 0.2 - 0.5$ and $x = 0$. Raw data are shown in fig. S2. In the upper four panels three peaks were found as predicted by the shell model. A fourth, lowest energy phonon in the bottom two panels, which disperses up from lower energies, is not expected from the shell model, but it appears far from the stretching peak. A more serious complication is that there appears to be a strong overlap between the stretching and bending phonons in the bottom two panels, which casts doubt on the accuracy of our fits. Further, scans in the (3, 3, 0) BZ do not yield more clear results at these $\mathbf{q}$ values. For this reasons we do not want to draw any conclusions for its dispersion for $x > 0.4$. Fig. S3 shows a direct comparison of background subtracted scans with each panel containing two scans measured at the same $\mathbf{q}$ but different $\mathbf{Q}$.

The results of our fits superimposed with the shell model prediction are shown in fig. S4. The bottom two branches are the bond-stretching branches, whereas the upper branch is the bond-bending one. Vertical bars represent the calculated $\mathbf{Q}$ dependent structure factors.

We discuss the bending branches first. There is a good qualitative agreement between the model and the experiment for the bending branches, except for the steeper upward dispersion of the lower branch. Another difference between the experiment and the model is a stronger intensity of the upper bending branch than what is calculated in the model. It appears that in the experiment the lower bending branch disperses more steeply than calculated and its character (eigenvector) is transferred to the upper branch.

The stretching branch disperses downwards generally following the calculations, except for the dip at near x=0.25, which is the interesting effect that we discuss in the letter.

5. Longitudinal Dispersion

Differences in the phonon dispersions between the transverse and longitudinal directions require different presentations of our results in order to better explain our analysis. Here we begin by discussing the calculated and experimental phonon dispersions in the two Brollouin zones.

Calculation results presented in fig. S5 show that there are two phonon branches in each zone with appreciable intensity. The bond-stretching branch is expected in both BZ, except it has a higher intensity in the (3, 3, 0) zone. Its dispersion deviates strongly from the experimental one. This deviation is discussed in the text of the letter. To add to that discussion, it is possible to add an extra force constant to the model to push down the dispersion (dashed line). We found that this artificial softening does not influence the phonon structure factor, thus the calculated structure factor structure factor may be used to assign the experimental stretching phonons though the calculated and experimental dispersions are very different.

Fortunately, the intense bending branches in the two zones are not the same, In the (3, 3, 0) zone the bending branch that disperses up from 31 meV at the zone center to 50 meV at the zone boundary has a strong intensity. In the (4, 4, 0) zone the strong intensity should be observed for the higher energy branch that disperses up from 42.5 meV. At the zone boundary both bending branches acquire nearly the same intensity. These predictions of the model are in an excellent agreement with experiment. To illustrate this point we show scans near the zone boundary at the wavevectors (3.45, 3.45, 0), (3.5, 3.5, 0), and (3.55, 3.55, 0) in fig. S6. It is clear that there are two peaks in the data whose frequencies correspond to the shell model prediction. What's more remarkable, is that the upper peak is suppressed on moving away from the zone boundary towards (3, 3, 0) and the opposite happens on going towards (4, -4, 0): This is exactly what the shell model predicts. These data show that the shell model is very reliable here and can be used as a guideline to separate the bond-bending from the bond-stretching peaks. Since the bending peaks dominate the scattering at the zone boundary, it is not possible to extract the bond stretching peaks here, however, according to the shell model, the intensity of the upper branch is nearly zero between (3, 3, 0) and (3.4, 3.4, 0), i.e. (2.6, 2.6, 0).

Fig. S7 shows the phonon data closer to the zone center superimposed in the two BZ. According to the model, the intensity of the upper peak in the (4, 4, 0) zone should be zero in the (3, 3, 0) zone. This is the most important feature of the lattice dynamics that enabled us to assign the bond-stretching mode as shown in fig. S6. For $\mathbf{q} < (h, h, 0)$, $h < 0.2$, the bond stretching peak is well separated from the bond-bending ones, so its assignment is unambiguous. The situation becomes less straightforward at $h \geq 0.2$, because there is an overlap between the low-energy side of the stretching peak and the upper bending peak. However, according to the shell model, the intensity of the upper bending peak seen in the (4, -4, 0) zone, is zero in the (3, 3, 0) zone, thus can we assign the intensity there to the stretching branch. The same applies up to $h = 0.4$; at higher h, the upper bending branch becomes intense even in the (3, 3, 0) zone, and it is impossible to separate the bending and stretching phonon peaks. This is the reason that we do not show the stretching peak frequencies for $h > 0.4$.

The upper bending and stretching peaks may hybridize away from the zone center and the zone boundary, which may affect result in an overestimate of the linewidth of the bond stretching peak. However comparison of the spectra in the (4, 4, 0) zone where the bond-bending phonons should dominate the scattering intensity (Fig. S5 right panel) with the (3, 3, 0) zone where the bond-stretching phonons should dominate (Fig. S5 left panel), does not appear to be the case. The bending peak seen in the (4, 4, 0) zone (red) and the stretching peak (black) in the (3, 3, 0) zone should lock on each other if there were hybridization, however, it is clear from fig. S7 that the two branches disperse independently.

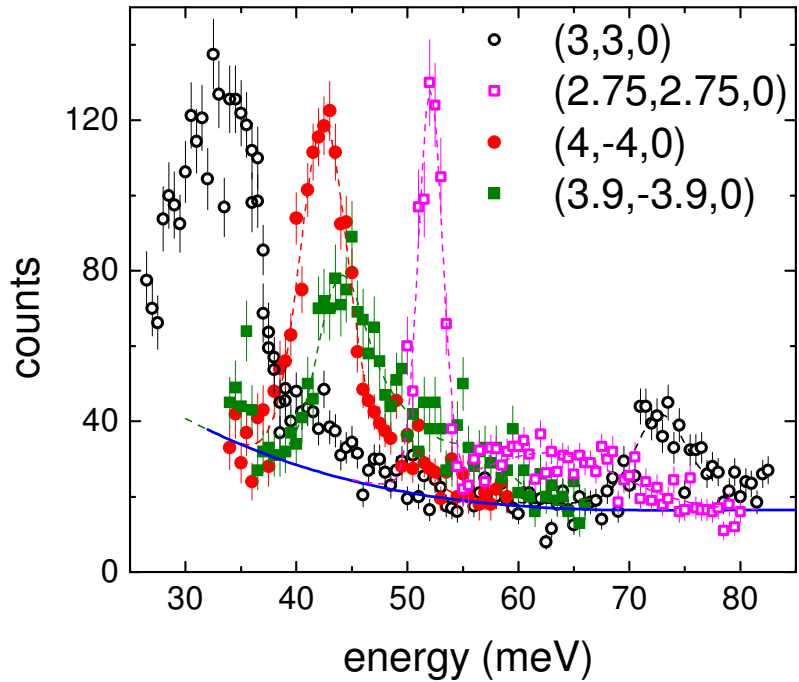

Figure S1: Background determination for longitudinal scans in (3,3,0) and (4,-4,0) Brillouin Zones. Dashed lines are guides to the eye. The blue line shows the estimated background function used for all longitudinal scans.

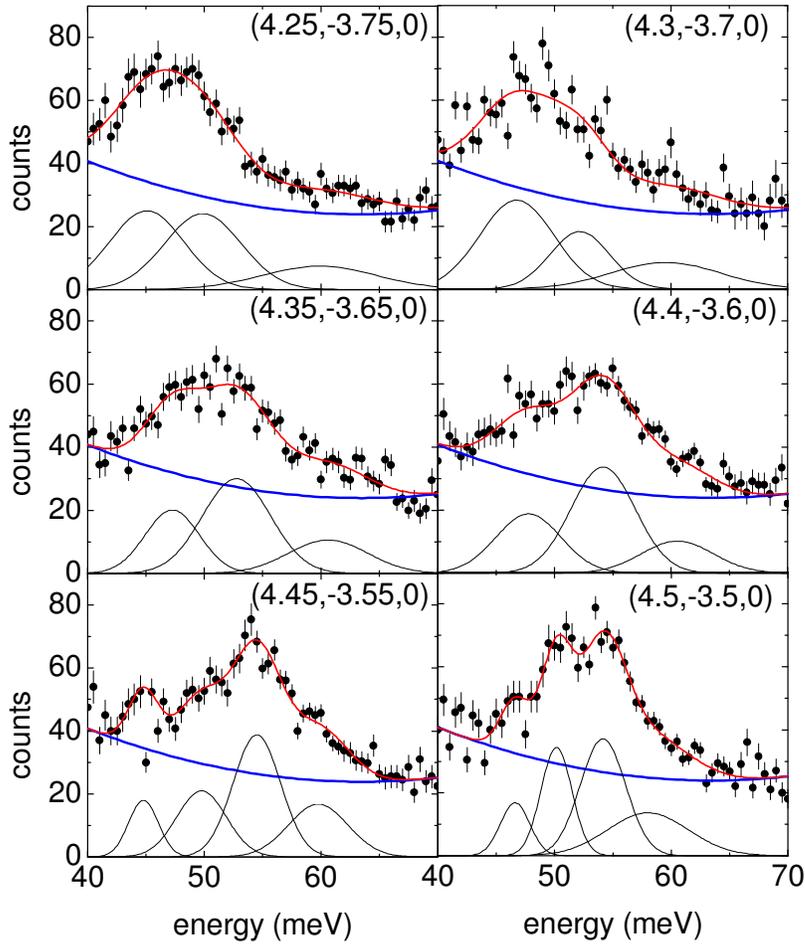

Figure S2: Transverse phonon scans at **Q** = (4+x, -4+x, 0), x = 0.25 – 0.5. Data points are experimental data. The blue line is a polynomial fit to the background data taken in part from the zone center scan at **Q** = (4, -4, 0) (blue data points in lowest right panel) and outer regions of the other scans. Black lines are gaussian fits to the data after subtracting the fitted polynom. Red lines are an overlay of all gaussians at a wave vector and the background function.

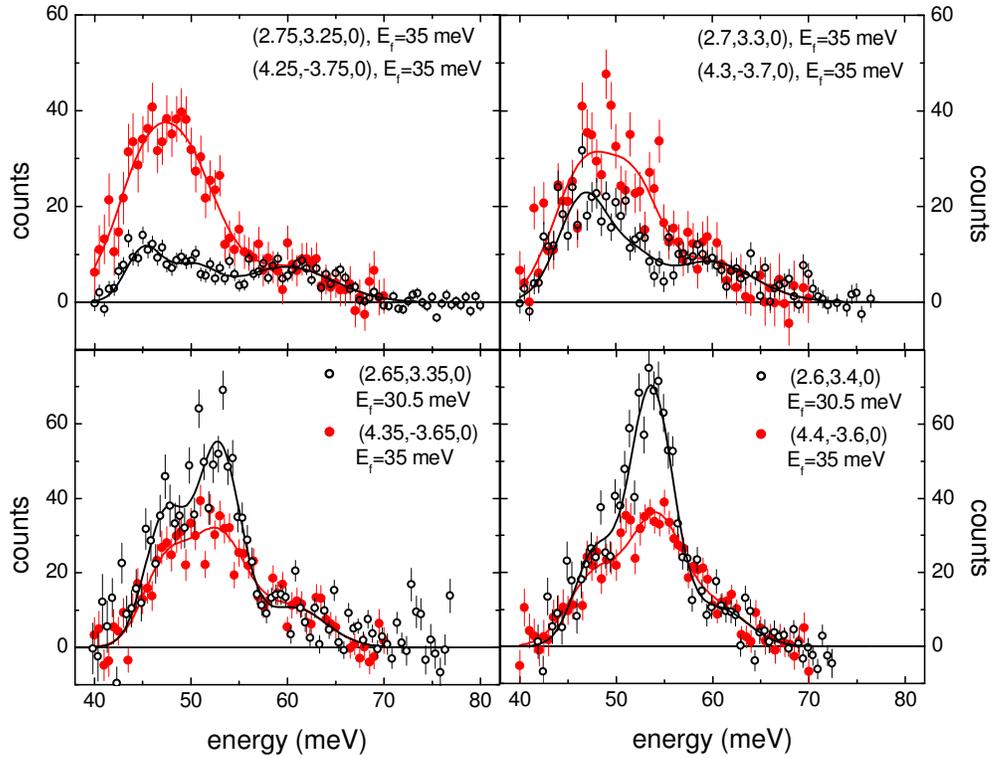

Figure S3: Superimposed background subtracted scans in the (3, 3, 0) and (4, 4, 0) Brillouin zones at reduced wave vectors **q** = (-h, h, 0), h = 0.25, 0.3, 0.35 and 0.4. Data in (3, 3, 0) BZ were scaled to have equal amplitude of the bond stretching gaussian fit as the (4, 4, 0) phonon scans for clarity. The gaussian fits representing the single phonons can be seen in figs. 2 and S2.

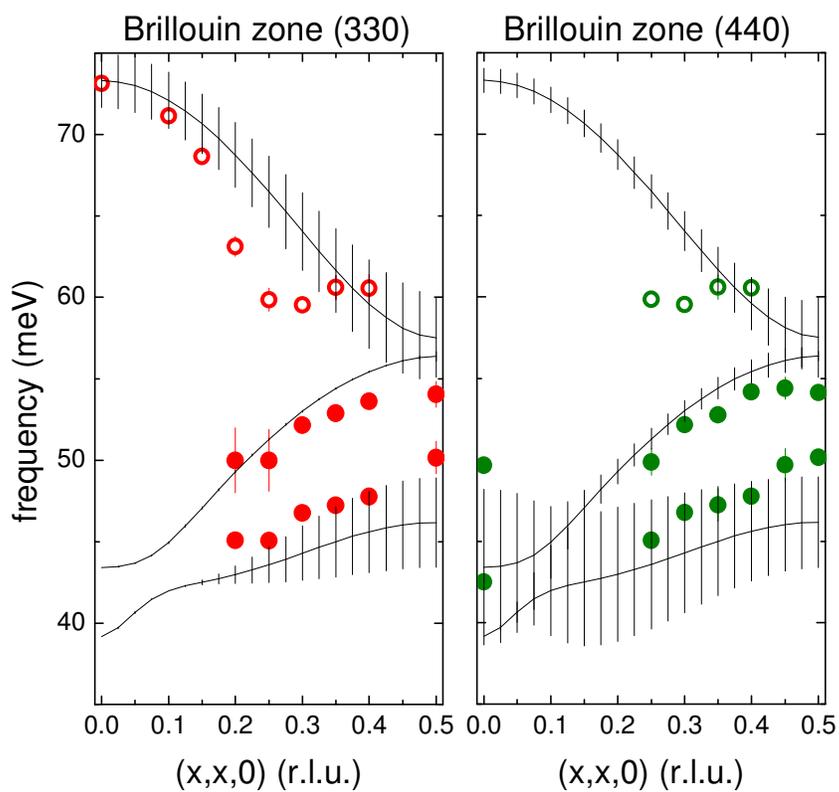

Figure S4: Dispersion of transverse phonon branches in the (110) direction in $La_{1.2}Sr_{1.8}Mn_2O_7$. Lines are calculated by a shell model. Vertical bars represent the calculated structure factors. Filled/open symbols represent phonon with bond bending/ stretching character. Diameter of the dots indicate the observed intensities of the phonon peaks.

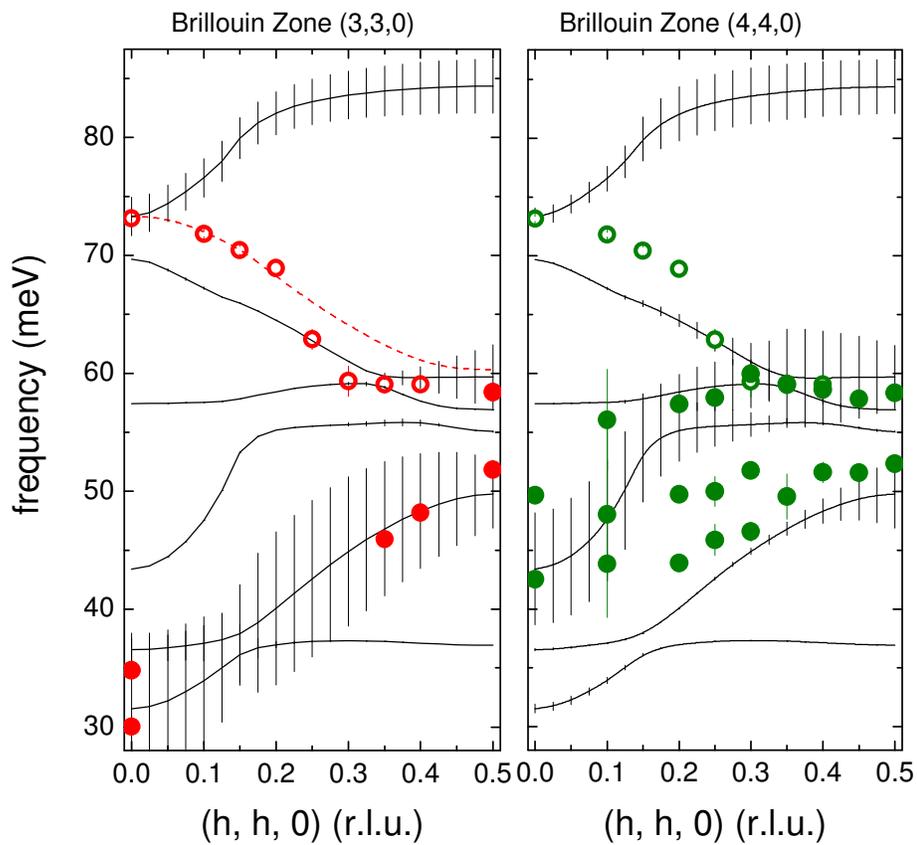

Figure S5: Dispersion of longitudinal phonon branches in the (110) direction measured in the (3, 3, 0) and (4, 4, 0) BZ. Lines are calculated by a shell model. Dashed line is calculated using a special force constant (see text). Filled/open symbols represent phonon with bond bending/ stretching character. Diameter of the dots indicate the observed intensities of the phonon peak.

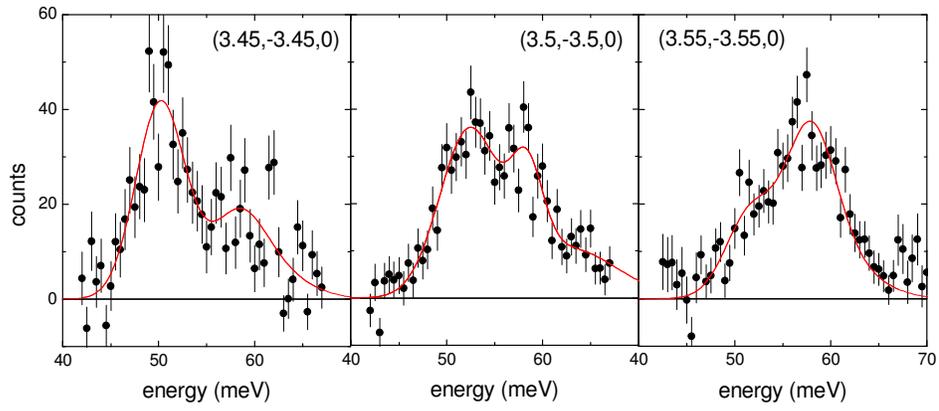

Figure S6: Background subtracted phonon scan at the zone boundary between the (3, 3, 0) and (4, 4, 0) zone centers and scans with $\Delta \mathbf{q} = (\pm 0.05, \pm 0.05, 0)$. Lines are superposition of gaussian fits for the phonon peaks.

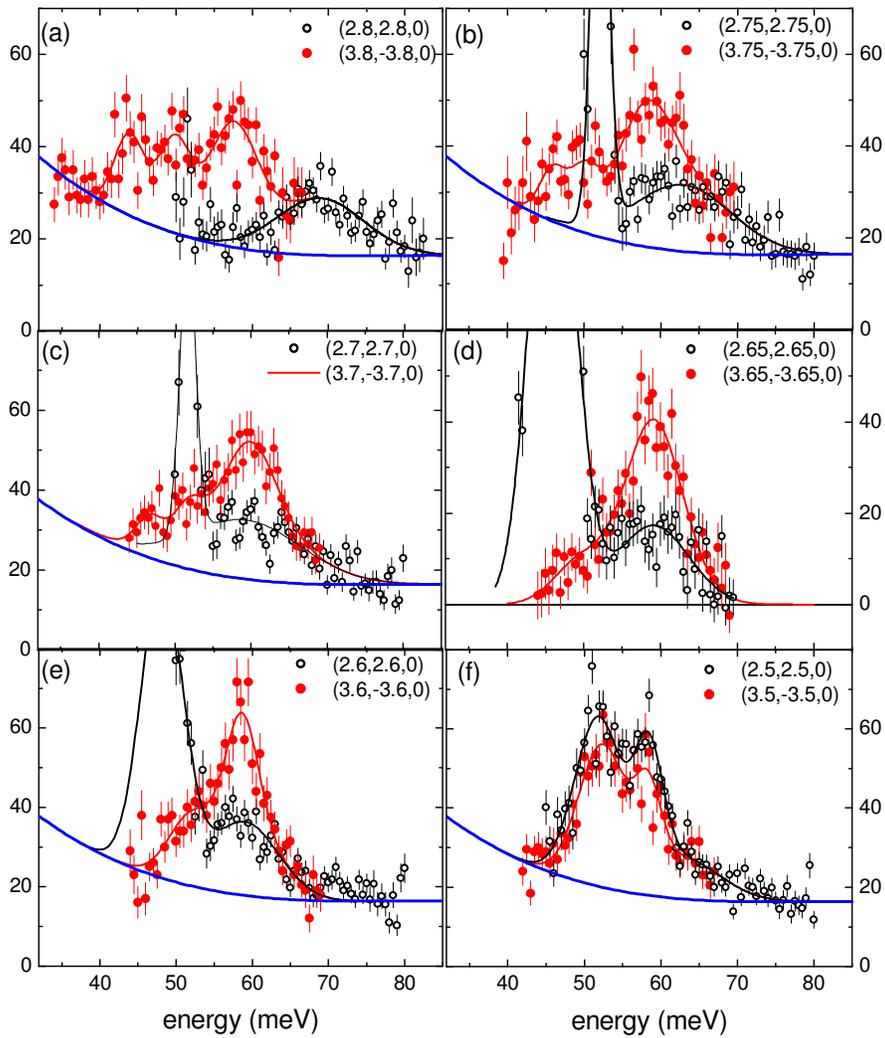

Figure S7: Raw data of longitudinal phonon scans in the (3, 3, 0) (open) and (4, -4, 0) (filled) BZ. The blue line in each panel denotes the background function as shown in fig. S1. The amplitudes of the gaussian fits for the bond stretching phonons in the (4, -4, 0) BZ are fixed to the values obtained for the scans in the (3, 3, 0) BZ.